\documentclass{article}

\usepackage[preprint]{neurips_2020}

\usepackage[utf8]{inputenc} 
\usepackage[T1]{fontenc}    
\usepackage{hyperref}       
\usepackage{url}            
\usepackage{booktabs}       
\usepackage{amsfonts}       
\usepackage{nicefrac}       
\usepackage{microtype}      

\usepackage{microtype}
\usepackage{graphicx}
\usepackage{subfigure}
\usepackage{booktabs} 
\usepackage{amsmath}  
\usepackage{nccmath}  
\usepackage{bbm}      
\usepackage{hyperref}
\usepackage{algorithm}
\usepackage{algorithmic}
\usepackage{caption}

\title{Estimating Causal Effects with the Neural Autoregressive Density Estimator}

\author{%
  Sergio Garrido \\
  Department of Transport\\
  Technical University of Denmark\\
  Lyngby, Denmark \\
  \texttt{shgm@dtu.dk} \\
  \And
  Stanislav Borysov \\
  Department of Transport\\
  Technical University of Denmark\\
  Lyngby, Denmark \\
  \texttt{stabo@dtu.dk} \\
  \And
  Jeppe Rich \\
  Department of Transport\\
  Technical University of Denmark\\
  Lyngby, Denmark \\
  \texttt{rich@dtu.dk} \\
  \And
  Francisco Pereira \\
  Department of Transport\\
  Technical University of Denmark\\
  Lyngby, Denmark \\
  \texttt{pereira@dtu.dk} \\
}

\begin{document}

\maketitle



\begin{abstract}
Estimation of causal effects is fundamental in situations were the underlying system will be subject to active interventions. Part of building a causal inference engine is defining how variables relate to each other, that is, defining the functional relationship between variables entailed by the graph conditional dependencies. In this paper, we deviate from the common assumption of linear relationships in causal models by making use of neural autoregressive density estimators and use them to estimate causal effects within Pearl's do-calculus framework. Using synthetic data, we show that the approach can retrieve causal effects from non-linear systems without explicitly modeling the interactions between the variables, and include confidence bands using the nonparametric bootstrap. We also explore scenarios that deviate from the ideal causal effect estimation setting such as poor data support or unobserved confounders. 

\end{abstract}


\section{Introduction}
\label{sec:Introduction}

One way of thinking about causal models is to consider them as inference engines \citep{pearl_transportability_2014}. These engines take causal assumptions and queries as input and give, as an output, answer to those queries. According to \cite{pearl_causality_2009, pearl_why_2018}, there exists a hierarchy, where the different levels in the hierarchy represent different types of causal queries that can be answered. In particular, a researcher might ask questions related to association, intervention, or counterfactual queries. Depending on the causal assumptions, different queries can be answered. This paper will focus on the second level of this hierarchy: namely, intervention queries.

The most important assumption required to compute interventional queries is that the topology of the causal graph is known. This topology can be elicited either by expert knowledge or by ``causal discovery" algorithms. Regardless of where the causal topology comes from, we will assume throughout the paper that the causal graph is known. However, sensitivity with respect to this assumption will be investigated in the experimental section.


Pearl's framework of causality was conceived in a non-parametric way, in the sense that he moved away from the assumption of cause-effect relations through simple parametric equations. These assumptions (e.g., a variable being a linear function of its causes) are often based on unrealistic simplifications and are not strictly required to estimate the models. While the assumption of linearity makes the estimation and, in some cases, the interpretation easier, such simplifications can have serious consequences for model parameters and predictions. In this paper, we build on recent advancements in the deep generative modeling literature  \citep{larochelle_nade_2011, goodfellow_dl_2016} and demonstrate how researchers can move away from these assumptions. To accommodate a flexible representation of cause and effect relationships we propose using functional forms based on neural networks to model such relationships. Particularly, we propose the use of neural autoregressive density estimators (NADE) \citep{larochelle_nade_2011, uria_rnade_2013} to model independent causal mechanisms \citep{scholkopf2019causality}. The contributions of this paper are the following:

\vspace{-\topsep} 
\begin{itemize}
    \item We propose a new way to estimate causal effects using Pearl's do-calculus. Using this framework, we can estimate arbitrary interventions as long as these can be identified. The 'back-door' criterion and the 'front-door' adjustment, explained in Section \ref{sec:simulations}, are special cases hereof.
    \item The proposed estimation method scales to high dimensions and complex relationships between the variables, as long as the structure representing these relationships is known.
    \item The resulting model preserves the properties of a generative model. It is therefore straightforward to create new data from the model or to impute missing values by doing ancestral sampling.
    \item Based on simulation we explore sensitivity with respect to different assumptions and demonstrate that the topology of the causal graph is indeed a sensitive assumption.  
    
\end{itemize}


\section{Related work}

Estimating causal effects using linear functions is the common practice in the social sciences literature \citep{angrist_ballot_1990, miguel_worms_2004, angrist_mhe_2008}. A common assumption in these studies is that the outcome variable is modelled as a linear function of the parameters and the covariates. An exception is when there is additional domain expertise as how the covariates relate with the outcome. This way of modeling causal effects is restrictive and mainly applies to the estimation of single causal effects. In such cases, non-linear functions of the covariates can be used. One of the main motivations of using neural networks in order to estimate causal mechanisms is that they can be viewed as universal approximators \cite{cybenko_approximator_1989}. It is worth noting that while neural networks are a natural choice due to it flexibility with respect to both continuous and discrete problems, other universal approximators or powerful conditional density estimators, such as Gaussian Processes could have been used in much the same way for continuous problems. 

\subsection{Non-linear causal effect estimation}

The idea of estimating causal effects in non-linear settings is not new. For example, \citet{hill_BART_2011} proposed a non-parametric Bayesian method---the Bayesian Additive Regression Trees (BART)---to estimate Average Treatment Effects (ATE). More recently, the literature on non-linear causal effect estimation has focused on estimating Individual Treatment Effects (ITE), for example, by using representations of a feed-forward neural network \citep{johansson_causalrepresentation_2016}, multi-task Gaussian processes \citep{alaa_gpcausal_2017}, or a variant of a Generative Adversarial Network \citep{yoon_ganite_2018}. The approach proposed in this paper can be viewed as a generalization of the idea in \citet{hartford_deepIV_2017}, where deep neural networks are used to estimate causal effects using instrumental variables (deepIV), and  further extended by \citet{bennett_deepGMM_IV_2019} to a Generalized Method of Moments approach for instrumental variables. However both papers are only concerned with the estimation of causal effects using instrumental variables. All of this literature is mainly focused on the potential outcome framework literature, which is not as general as Pearl's causal framework and where the assumptions related to the causal graph are not stated as transparently as in Pearl's framework. 

\subsection{Neural autorregresive density estimators}

The literature related to autoregressive density estimators is comprehensive. As mentioned in the Introduction, we are interested in approaches that use neural networks to learn useful representations from data. Here, we can highlight \citet{bengio_ARNN_1999} who introduced the idea of using neural networks to estimate the parameters of distributions of any variable given previously observed variables. \citet{larochelle_nade_2011, uria_rnade_2013, germain_made_2015} extended this idea in order to, for example, make the neural network training more efficient, or model a richer family of distributions. To the best of our knowledge, neural autoregressive density estimators have not been used for causal inference.

\section{Methods}

Consider a set of $J$ random variables $X_1,...,X_J$ representing a system of interest. Further, let lowercase letter variables $x_1,...,x_J$ denote realizations of these random variables. Mathematically, the interventional distribution of an outcome variable $X_{o}$ under an intervention $X_{j}=x_{j}$ can be defined as $P(X_{o} \mid do(X_{j}=x_{j}))$ or $P_{do(X_{j}=x_{j})}(X_{o})$ \citep{pearl_causality_2009}. In these equations, the $do$ operator, ``$do(X_{j}=x_{j})$'', expresses that the variable $X_{j}$ is forced to be $x_{j}$. This distribution allows us to estimate other quantities of interest, for example, the Average Treatment Effect (ATE) of treatment $x_{j}$ relative to treatment $\hat{x}_{j}$, defined as
\begin{ceqn}\label{eq:ate}
    \begin{align}
        ATE = E[X_{o} \mid do(X_{j}=x_{j})]-E[X_{o} \mid do(X_{j}=\hat{x_{j}})].
    \end{align}
\end{ceqn}

\subsection{Assumptions}
\label{ssec:assumptions}
Since this paper deals with the estimation of causal models, the following assumptions are required:
\begin{itemize}
    \item The underlying causal graph of the process is known. This graph is a Directed Acyclical Graph (DAG). Relaxations with respect to this assumption are explored in Section~\ref{ssec: unobserved_confounder_experiment}. 
    \item The distribution family of every variable in the process is known. This is not so restrictive as we will see in Section \ref{sec:cont_conf}.
\end{itemize}

These assumptions, either implicitly or explicitly, are in accordance with the common practice when estimating interventional distributions. However, it is worth stressing that in contrast to the majority of models in the literature, we do not make any assumptions about the functional form of the relationships between the variables and the corresponding parameters. These assumptions are relaxed exactly for the reason of being able to estimate more robust causal effect models for which these relationships are not known in advance. In particular, it is interesting to consider how the performance of the approximation depend on the data support.

\subsection{Autoregressive density estimators}
\label{ssect: autoregressive_density_estimators}

Before introducing the relationship with causal models, we briefly explain the concept of neural autoregressive density estimators. These models are a family of generative models that attempt to estimate the joint distribution of the variables by factorizing the joint distribution into a product of conditional distributions \citep{bengio_ARNN_1999, larochelle_nade_2011, germain_made_2015}. By using the chain rule of probability, any joint probability distribution $P(X)$ can be factorized as a conditional distribution where the probability of $X_j$ is conditioned on a function $f_{j}$ of its ``parents" ($\mathrm{PA}(X_{j})$). This applies to any arbitrary graph that contains all the variables of interest:
\begin{ceqn}
    \begin{align} \label{eq:markov_factorization}
        P(X) = \prod_{j}^{J} P(X_{j} \mid f_{j}(\mathrm{PA}(X_{j}))).
    \end{align}
\end{ceqn}
Even though \cite{bengio_ARNN_1999} suggest that these models can be constructed on the basis of probabilistic causal models, most of the machine learning literature has moved away from this approach. Instead, the machine learning literature has focused on proving that the densities estimated by these models are robust with respect to changes in the joint probability factorization. There is a body of evidence from the deep generative model literature that suggests this is indeed possible \citep{uria_rnade_2013, germain_made_2015, papamakarios_maf_2017}. Some of these models are valid as an approximation of associational inference. However, as we are interested in causal inference, and more specifically, interventional type of inference, further assumptions with respect to the factorization of the distribution are required. In other words, we cannot just impose an arbitrary factorization of conditional distributions as in the body of neural autoregressive estimator literature. Instead, we must apply a ``causal factorization'' as discussed in \cite{parascandolo_learningICM_2017}. The individual conditional distributions are also called Individual Causal Mechanisms (ICM) or Markov Kernels \citep{peters_elements_2017}.

\begin{figure}[!htb]
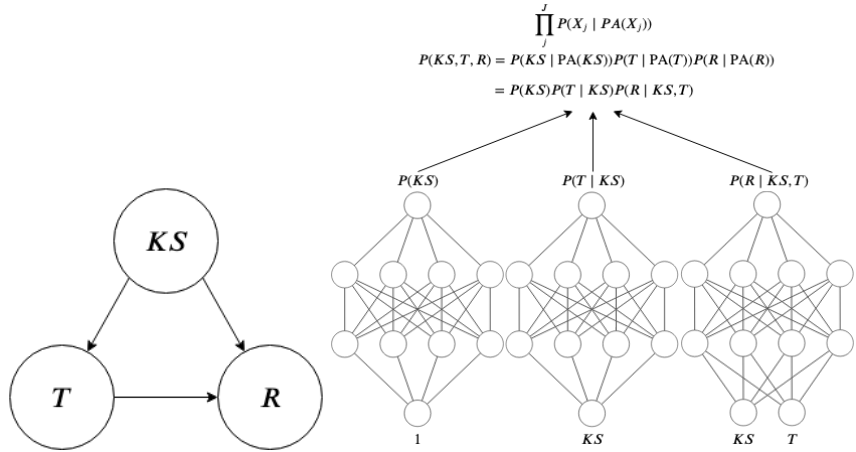

  \centering
  \includegraphics[width=0.3\textwidth]{assets/kidney_graph.png}
  \includegraphics[width=0.5\textwidth]{assets/causal_network.png}
  \caption{Example of a neural autoregressive density estimator model. (Left) The causal graph we want to represent with neural networks. (Right) The neural autoregressive density estimator representing the causal graph. All the variables are parameterized as Bernoulli distributed random variables.}
  \label{fig:neural_model}
\end{figure}

We propose to parametrize the functions $f_{j}$ as independent fully connected feed-forward neural networks, and use them as independent causal mechanisms in order to estimate the effects of interventions. For every variable $X_{j}$ of our system there is a neural network. Each of the neural networks takes the parents of the variable of interest $\mathrm{PA}(X_{j})$ and outputs parameters that define the distribution of $X_{j}$. For example, if $X_{j}$ is a binary variable, the network outputs a probability, $p$, of $X_{j}=1$. Likewise, if $X_{j}$ is continuous and approximately Gaussian distributed, the neural network would output the mean $\mu$ and the standard deviation $\sigma$ of $X_{j}$. These networks are trained jointly, using the negative log-likelihood in Equation~\eqref{eq:markov_factorization} as the loss function. Figure~\ref{fig:neural_model} illustrate an example of such a model for a causal graph of the ``kidney stone'' example \citep{charig_kidney_1986, peters_elements_2017}. This graph contains assumptions related to the conditional dependencies between variables, for example, that $R$ (recovery) is dependent of both $KS$ (size of kidney stones) and $T$ (treatments). Each network takes parents of each variable as input and outputs the parameters of the respective variable. In this case, the output is a single neuron, since each random variable follows a Bernoulli distribution with parameter $p$.
  
\subsection{Connection to Pearl's Do-Calculus}
\label{ssect: connect_to_do_calc}

The way to connect the model described in the previous subsection to the do-calculus framework is by realizing that Equation~\eqref{eq:markov_factorization} is closely related to the truncated factorization formula \citep{pearl_causality_2009}. For an intervention where $X_{j}$ takes a single value, $do(X_{j}=x'_{j})$, this corresponds to
\begin{multline}\label{eq:truncated_factorization}
        P(X_{o \neq j} \mid do(X_{j}=x'_{j})) = 
        \begin{cases}
            \prod_{o \neq j} P(X_{o} \mid \mathrm{PA}(X_{o})) &\text{if } X_{j}=x'_{j} \\
            0 &\text{otherwise}
        \end{cases}
\end{multline}
The neural networks will approximate each of these conditional distributions or independent causal mechanisms \citep{peters_elements_2017} that are fundamental to the estimation of the causal effects. In order to compute causal effects, it is required that effects can be identified and that we use the rules of the do-calculus as explained in more detail in \cite{pearl_causality_2009, pearl_do_2012} and \cite{pearl_why_2018}.


\section{Simulation experiments}
\label{sec:simulations}

In order to present a use-case of the neural network approximation of the independent causal mechanisms we present several simulation experiments based on the ``kidney stones" example \citep{charig_kidney_1986, peters_elements_2017}. The case, which has been studied widely in the literature to show the difference between conditioning and intervening, is an excellent reference case for our approach.

In total, we run nine different experiments. In the first two, we show that our approach is able to recover linear functions both in binary and continuous variable settings. In the following seven experiments, we illustrate the ability of the approach to recover causal effects in non-linear settings where, as expected, a linear estimator is unable to recover the true causal effects. In addition, it is shown that the approach can be extended to the estimation of interventions in causal graphs where a ``front-door adjustment" is required. This contrasts with the literature of using neural networks to estimate causal effects, restricted to specific scenarios such as instrumental variables \citep{hartford_deepIV_2017}.  A neural network with no activation functions is used as a benchmark. Such neural network is equivalent to a linear regression model. More details on the experimental setup can be found in the Appendix~\ref{apx:experimental_setup}.

The kidney stone example contains three variables, stone size ($KS$), treatment type, ($T$), and Recovery ($R$). The assumed causal graph is presented in Figure~\ref{fig:neural_model}, and we assume we observe all the variables, unless stated otherwise. The original observed values from \citet{charig_kidney_1986} are reproduced in Table~\ref{tab:original_data} while the Data Generation Process (DGP) is presented in Algorithm~\ref{alg:binary_model}.

\begin{table}[!htb]
  \centering
  \vskip 0.15in
  \caption{Distribution of the original kidney stone example}
  \begin{tabular}{lll}
        \toprule
        \multicolumn{3}{c}{Treatment type}          \\
        \cmidrule(r){2-3}
        Size     &           A             &        B       \\
        \midrule
        Small    & \textbf{93\%} (81/87)    & 87\% (234/270)  \\
        Large    & \textbf{73\%} (192/263)  & 69\% (55/80)    \\
        \midrule
        Total    & 78\% (273/350)          & \textbf{83\%} (289/350)  \\
        \bottomrule
    \end{tabular}
  \label{tab:original_data}
  \vskip 0.1in
\end{table}

\begin{algorithm}
    \caption{DGP for the kidney stone example}
    \label{alg:binary_model}
    \begin{algorithmic}[1]
        \FOR{each individual $i$}
        \STATE Draw $KS \sim P(KS=Small)=0.51 \: (357/700)$ \label{alg:line:confounder}
        \STATE Draw $T \sim P(T=A \mid KS=S)=0.24 \: (87/357) \: \textrm{and} \newline 
        \: P(T=A \mid KS=L)=0.77 \: (263/343)$
        \STATE Draw $R \: \textrm{with} \: P(R \mid KS, T) \: \textrm{as in Table~\ref{tab:original_data}.}$ \label{alg:line:recovery}
        \ENDFOR
    \end{algorithmic}
\end{algorithm}

The distribution of the recovery, $R$, under an intervention for a treatment $T$, is defined as $P(R \mid do(T=t))$, and is calculated analytically using a result known as the ``backdoor adjustment formula'' \citep{pearl_causality_2009}. This result can be derived by using the rules of do-calculus in combination with the estimated probability distributions:
\begin{multline}\label{eq:backdoor_adjustement}
    P(R=1 \mid do(T=A)) = 
        \sum_{ks} P(R=1 \mid T=A, KS=ks)P(KS=ks) \\ 
    P(R=1 \mid do(T=B)) = 
        \sum_{ks} P(R=1 \mid T=B, KS=ks)P(KS=ks) \\ 
    ATE                 = 
        E[R=1 \mid do(T=A)]-E[R=1 \mid do(T=B)]. 
\end{multline}

In a setting where $KS$ is continuous, the sum is replaced by an integral.

\subsection{Experiment I: Binary Variables}

To estimate causal effects with binary outcomes from Algorithm~\ref{alg:binary_model}, we need to estimate the distributions in Equation~\ref{eq:backdoor_adjustement}. In order to do this, we used neural networks as shown in Figure~\ref{fig:neural_model}. There is one neural network for each variable in the system. Each neural network takes the parents of the corresponding variable as input and produces the parameters of a distribution as output. In this case, all the variables were modelled as random variables with Bernoulli distributions. This means that, for all neural networks in our system, the output of each one of them was a single parameter $p$.

Table \ref{tab:binary_model_results} presents the estimated conditionals. The fact that we are able to recover the probabilities from the generated data is not surprising, since knowing the structure of the causal graph reduces the estimation problem to a conditional density estimation problem. For a linear model, the resulting ATE is 6.43\%. This value is close to the analytical one since, in this model, the probability of the outcome does not include non-linear relationships between the covariates.

\begin{table}[!htb]
  \centering
  \vskip 0.15in
  \caption{Estimates of the probabilities estimated by the proposed model, and Average Treatment effect in the Kidney stone example with data generated only from binary variables. The Interventional distributions were estimated using Equation~\ref{eq:backdoor_adjustement}}
  \begin{tabular}{ll}
    \toprule
    \multicolumn{2}{c}{Neural model results} \\
    \midrule
    $p(large)$                   & 48.97\%  \\
    $p(R = 1 \mid large, A)$     & 75.58\%  \\
    $p(R = 1 \mid large, B)$     & 69.68\%  \\
    $p(R = 1 \mid small, A)$     & 95.68\%  \\
    $p(R = 1 \mid small, B)$     & 88.94\%  \\
    \midrule
    Neural ATE  & 6.32\% \\
    Linear ATE  & 6.43\% \\
    True ATE  & 5.3\% \\
    \bottomrule
  \end{tabular}
  \label{tab:binary_model_results}
  \vskip 0.1in
\end{table}

\subsection{Experiment II: Continuous response variables}

In the second simulation experiment, we generated the confounding variable $KS$, and the treatment variable $T$ as before, but now changed the recovery variable $R$ to be normally distributed with $R \sim N (R \mid \mu = T*4 + exp(KS), \sigma=2)$. That is, we changed Line~\ref{alg:line:recovery} of Algorithm~\ref{alg:binary_model} to follow a Normal distribution conditional on treatment and kidney size. The analytical causal effect of the treatment variable is 4. In this case, the output of the neural network that corresponds to the recovery variable has two units, one for the mean and one for the variance of a normal distribution. In that way, we are able to estimate the likelihood of the parameters, conditioned on our data, given the assumption of a normally distributed recovery variable.  Table~\ref{tab:continuous_case} shows the results of the proposed method. The treatment effect from the generated data is equivalent to the effect derived analytically. The difference between the analytical ATE and the neural approach are 0.11 percentage points, while the difference with the linear model are 27 basis points. Neither of these differences is large enough to invalidate one of the approaches. However, this phenomenon does tell that the neural approach is competitive also in linear settings.

\begin{table}[!htb]
  \centering
  \caption{Estimates of the ATE under a continuous recovery model.}
  \label{tab:continuous_case}
  \vskip 0.15in
  \begin{tabular}{ll}
    \toprule
    \multicolumn{2}{c}{Continuous case ATE (Equation~ \eqref{eq:backdoor_adjustement})} \\
    \midrule
    Analytical ATE & 4    \\
    Neural ATE     & 3.89 \\
    Linear ATE     & 3.73 \\
    \bottomrule
  \end{tabular}
  \vskip -0.1in
\end{table}

\subsection{Experiment III: Continuous confounding variables}
\label{sec:cont_conf}

In the third simulation experiment, the confounding variable $KS$ is generated as a continuous variable. In this experiment, the aim is to test the impact of the second assumption in Section~\ref{ssec:assumptions} that the distribution family of every variable in the process is known. To do this, two different data sets with different distributions of the confounding variable---gamma and log-normal are generated. That is, we changed Line~\ref{alg:line:confounder} in Algorithm~\ref{alg:binary_model} with a gamma distribution (left column of Table~\ref{tab:continuous_case_lognormal_gamma}) or a log-normal distribution (right column of the same table). Due to the non-differentiable nature of its parameters, gamma distributions cannot be learnt in a straightforward way using neural networks. As a result, we used a log-normal parametrization for both generated data sets. That is, in the first neural network of Figure~\ref{fig:neural_model}, the output is now a mean and a variance with which we can evaluate the likelihood using the kidney stone data.

The treatment variable is still a binary variable. A cutoff equal to $10$ (the mean of the confounder) is applied to decide whether the kidney stone is small or large $T=\mathbbm{1}\{KS>10\}$. After deciding whether the kidney stone is large or small, we assigned the treatment using the same probabilities as in Table~\ref{tab:original_data}. Finally, the response variable is defined as a normal distribution $R \sim N (R \mid \mu=T*4 + KS,\sigma=2)$. In the second data set, the confounding variable is drawn from a log-normal distribution with $\mu=2.5$ and $\sigma=0.25$. The rest of the Data Generation Process (DGP) is the same. 

The DGP for this experiment can be summarized as follows:
\begin{algorithm}
    \caption{DGP for a continuous confounder}
    \label{alg:continuous_outcome}
    \begin{algorithmic}[1]
      \FOR{each individual $i$}
      \STATE Draw $KS \sim$ Gamma$(5, 2)$ or, in the second set $KS \sim$ Log-Normal$(2.5, 0.25)$
      \IF{$KS>10$}
        \STATE Draw $T$ with $P(T=A) = 263/343$
      \ELSIF{$KS \leq 10$}
        \STATE Draw $T$ with $P(T=A) = 87/357$
      \ENDIF
      \STATE Draw $R \sim N (R \mid \mu=T*4 + KS,\sigma=2)$
      \ENDFOR
    \end{algorithmic}
\end{algorithm}

Since we have a continuous confounding variable, the sum in Equation~\eqref{eq:backdoor_adjustement} turns into an integral. If the integral is mathematically intractable, numerical approximations can be used. It is here proposed to use Monte Carlo integration in order to obtain an approximation of the interventional distribution and the ATE. To do this, we sample from the inferred distribution and propagate forward through the neural network to get different estimates of the response variable. 

The results of this experiment are shown in Table~\ref{tab:continuous_case_lognormal_gamma}. Even though the analytical treatment effects are not recovered perfectly, they are approximated quite accurately in this way. The ATE is the same for any number of samples, as the functional relation between the mean of the outcome and the confounder is linear.

These results also reveal that the assumption of knowing the exact parametrization of the distribution is not critical. This is supported by the possibility of increasing the complexity of any modeled variable by using a mixture of densities \citep{bishop_mdn_1994} or normalizing flows \citep{agnelli_cnf_2010, tabak_nf_2013, rezende_nf_2015}. In other words, we can increase the number of output units of any neural network to parametrize a variable with a highly complex distribution. Since the DGP of the outcome variable is linear, the linear benchmark is able to recover the treatment effects in the same way as the proposed approach.

\begin{table}[!ht]
  \centering
  \caption{Estimates of the ATE with a continuous confounding variable generated by a Gamma and a Log-normal distribution.}
  \label{tab:continuous_case_lognormal_gamma}
  \vskip 0.15in
  \begin{tabular}{l c c}
    \toprule
    \multicolumn{3}{c}{Continuous confounding ATE (Equation. \eqref{eq:backdoor_adjustement})} \\
    \midrule
     & Gamma confounding & Log-normal confounding \\
    \midrule
    1 sample        & 3.77 & 4.15 \\
    5 samples       & 3.77 & 4.15 \\
    25 samples      & 3.77 & 4.15 \\
    50 samples      & 3.77 & 4.15 \\
    \midrule
    \multicolumn{2}{c}{Total analytical ATE} & 4 \\
    \bottomrule
  \end{tabular}
  \vskip -0.1in
\end{table}

\subsection{Experiment IV: Non-linear relation between all the variables in the system}
\label{sssec:non-linear case}

In the fourth simulation study, data for both the confounding and the recovery value are generated as continuous variables. In this case, the effect of the confounding and treatment variable on the recovery is non-linear. In fact, this setting allows us to compute treatment effects conditioned on the confounding variable.

The confounding variable was simulated from a Log-normal distribution, $KS \sim$ Log-Normal$(\mu=2.5, \sigma=0.25)$. The treatment, as in the previous case, was modeled using a Bernoulli distribution with probability $p=\frac{1}{1+exp((KS-\mu)/10)}$. $p$ was modeled using a transformation of $KS$ in order to avoid extreme probabilities. Finally, the response variable was simulated from a normal distribution, $R \sim N \big(\frac{50T}{KS+3}, 1\big)$. 

We summarize the DGP in the following algorithm:

\begin{algorithm}
    \caption{DGP for the non-linear relation experiment}
    \label{alg:non_linear}
    \begin{algorithmic}[1]
        \FOR{each individual $i$}
        \STATE Draw $KS \sim$ Log-Normal$(2.5, 0.25)$, we subtract the mean of $KS$.
        \STATE Draw $T$ with $P(T=A) =  \frac{1}{1+exp(-KS/10)}$
        \STATE Draw $R \sim N \big(\frac{50T}{KS+3}, 1\big)$
        \ENDFOR
    \end{algorithmic}
\end{algorithm}

\begin{figure*}[!htb]
  \includegraphics[width=0.49\textwidth]{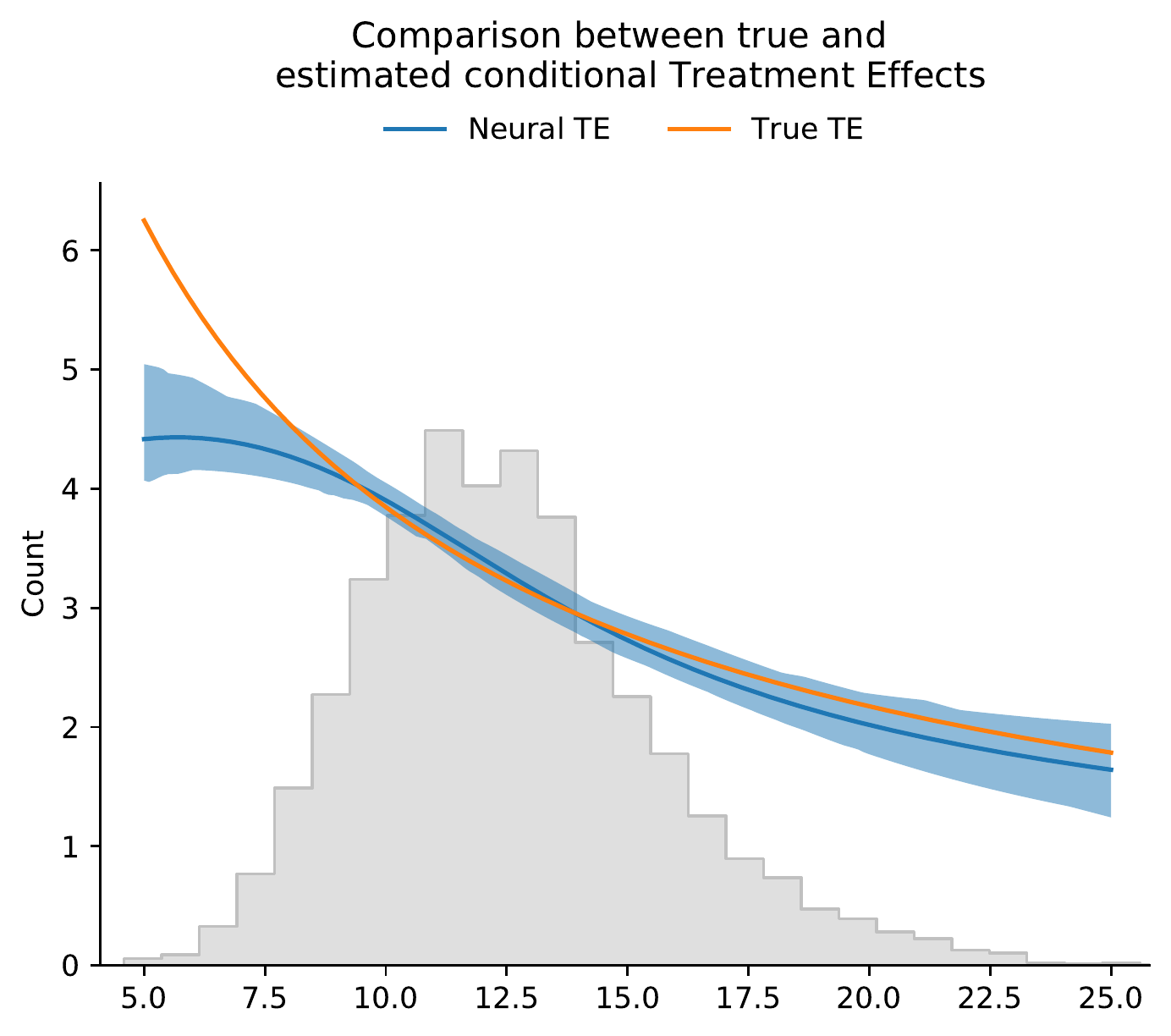}
  \includegraphics[width=0.49\textwidth]{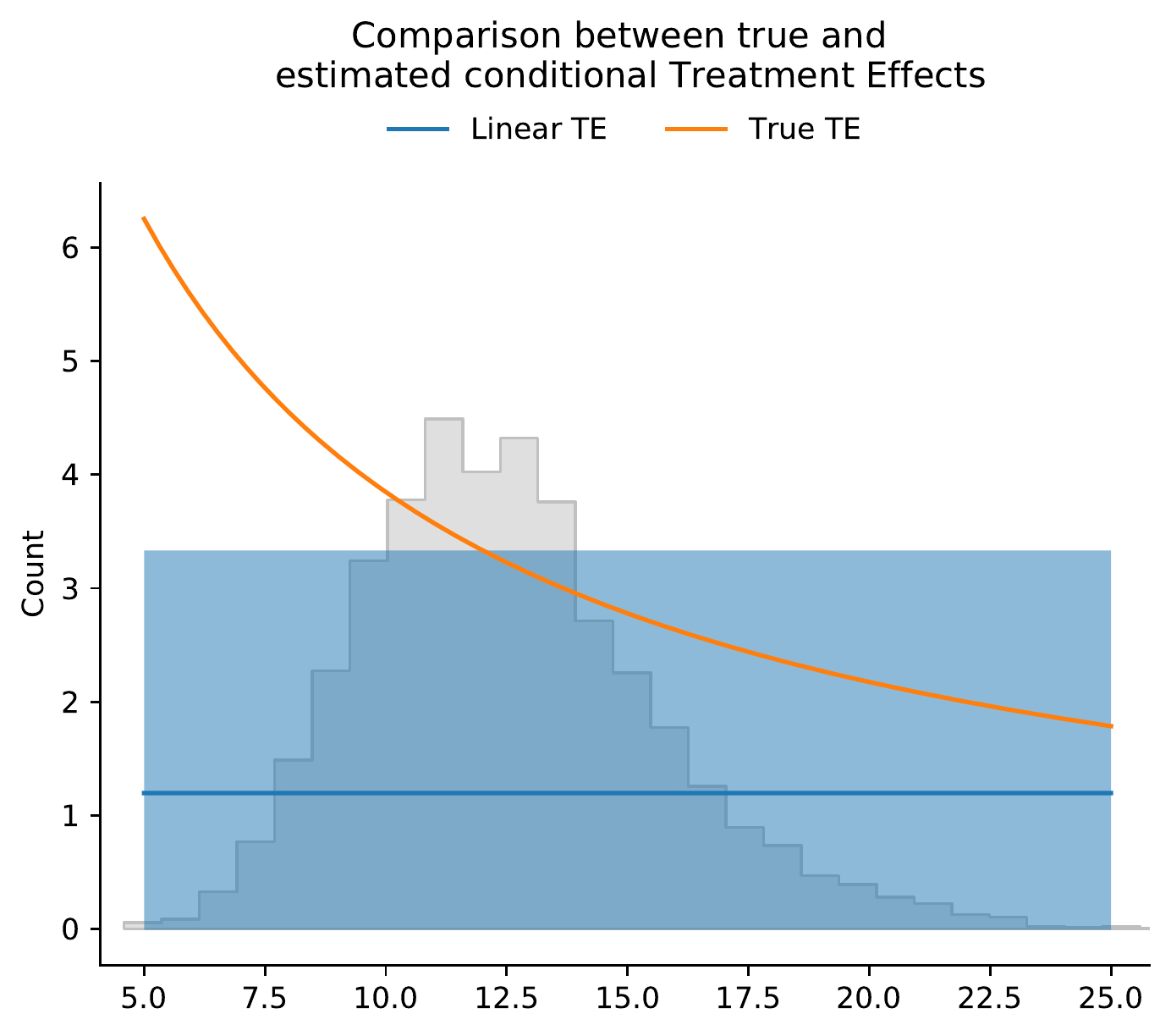}
  \caption{Comparison between the linear and neural network models using the back-door adjustment formula. The solid blue line represents the mean of the bootstrap, while the bands around the mean represent the 90\% confidence bands. A histogram of $KS$ is represented in the background of the plots.}
  \label{fig:non-linear_comparison}
\end{figure*}

These definitions make treatment effects non-linear, as illustrated in Figure~\ref{fig:non-linear_comparison}. Where the means, and 90\% confidence bands using the nonparametric bootstrap \citep{efron2016computer} are plotted for both the neural approach (left) and the linear approach (right).  The results are not surprising for the linear model, as the linear model is only able to retrieve a pooled estimate of the treatment effects. On the other hand, the neural network model is able to estimate the individual treatment precisely. 

Nevertheless, Figure~\ref{fig:non-linear_comparison} indicate some weakness in our approach. As we get further away from the support of the data, the estimated treatment effects gets less precise. For values below $KS=8$ the estimated treatment effect deviates drastically from the analytical one. The possibility of accurately estimating conditional treatment effects using neural networks depends heavily on the ability of the neural network to generalize beyond the support of the data.

Using a linear model, the ATE is constant. That is, regardless of the value of $KS$, a unique value for the causal effect is found. The problem with this estimation is that the non-linear relation between the confounding variable and the treatment is aggregated giving rise to aggregation bias \cite{palgrave_aggregation_2016}. This bias can be of any order of magnitude. For example, in our simulation study, with $R \sim N (R \mid 4*T*exp(-99l) + KS)$, the aggregated treatment effect would be close to 2, underestimating the effect for small $KS$ and overestimating it for large values.

\subsection{Experiment V: The front door adjustment}

\begin{figure}
    \centering
    \includegraphics[width=0.3\textwidth]{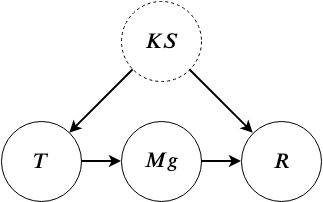}
    \caption{Graphical model of an extended kidney stone example where the confounder $KS$ is not observed.}
    \label{fig:front-door}
\end{figure}

In the previous simulation experiments all the variables in the causal graph were observed. In fact, because of the simplicity of the graph, the causal effects collapsed to a conditional estimation as noted in Section~\ref{sssec:non-linear case}. In this simulation study, the aim is to highlight that our approach works in conditions where there are unobserved confounders and, as a result, the back-door criterion does not hold. In Figure~\ref{fig:front-door} a causal graph with those conditions is presented. The dashed line of the $KS$ circle denotes an unobserved variable and, hence, Equation~\ref{eq:backdoor_adjustement} cannot be used.

In this case, $KS$ was simulated from a standard normal distribution, and $T$ was simulated from a normal distribution $T \sim N (\sin(KS), 0.1)$. Moreover, $Mg$ which is a mediator variable representing the amount of a certain chemical (in this case Magnesium) in the treatment, was simulated from a normal distribution $Mg \sim N (1+T^{2}, 0.1)$, while $R$ was simulated from a normal distribution $R \sim N \Big(\sin(KS^{2})+\frac{5}{Mg}, 0.1\Big)$. The graphical model and the generative process are shown in Figure~\ref{fig:front-door}.

\begin{algorithm}
    \caption{Generative process for the front-door adjustment study}
    \label{alg:front_door}
    \begin{algorithmic}[1]
        \FOR{each individual $i$}
        \STATE Draw $KS \sim N(0, 1)$
        \STATE Draw $T \sim N(\sin(KS), 0.1)$
        \STATE Draw $Mg \sim N(1+T^{2}, 0.1)$
        \STATE Draw $R \sim N \Big(\sin(KS^{2})+\frac{5}{Mg}, 0.1\Big)$
        \ENDFOR
    \end{algorithmic}
\end{algorithm}

Given the conditions of the causal graph, the back-door criterion cannot be used, and we must rely on other methods to estimate the causal effect of $T$ on $R$. The method to identify the causal effect in this graph is called the front-door adjustment \citep{pearl_causality_2009}
\begin{multline}\label{eq:frontdoor_adjustment}
    P(R \mid do(T=\hat{t})) = \\
    \int_{mg} P(Mg=mg \mid T=\hat{t}) 
    \int_{t'} P(R \mid Mg=mg, T=t') P(T=t').
\end{multline}

\subsubsection{Auxiliary networks}
As can be seen from Equation~(\ref{eq:frontdoor_adjustment}), the neural autoregressive model does not provide all the conditional distributions necessary for the estimation of the treatment effects. In this case, we have to estimate an ``auxiliary network'' which can be used to estimate the causal effects. For our particular causal graph, and the causal effect identified in Equation~\ref{eq:frontdoor_adjustment}, the neural network to be estimated takes $Mg$ and $T$ as input and predicts parameters associated to the distribution of $R$. With that auxiliary network, it is possible to estimate the causal effects.

\subsubsection{Results}
Since the analytical solution of this causal graph is not readily available, the true causal effects were simulated from binned conditional distributions using the true generative process. We compared the distributions using two values of $T=0$ and $T=0.5$ ($T$ is constrained between -1 and 1). Furthermore, the integrals in Equation~(\ref{eq:frontdoor_adjustment}) are approximated using Monte Carlo integration as explained in Section~(\ref{sec:cont_conf}).

\begin{figure*}[!htb]
  \includegraphics[width=0.3\textwidth]{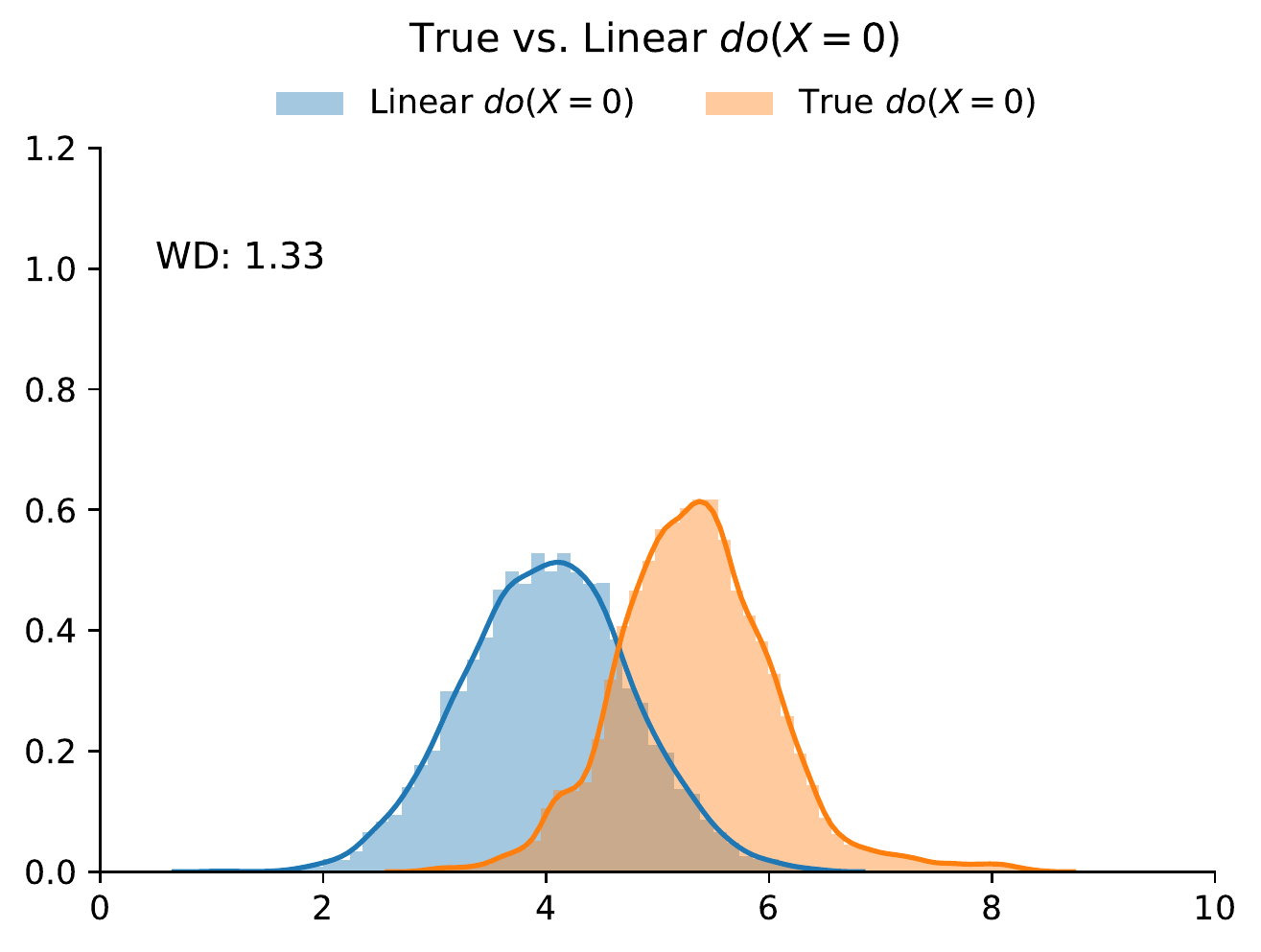}
  \includegraphics[width=0.3\textwidth]{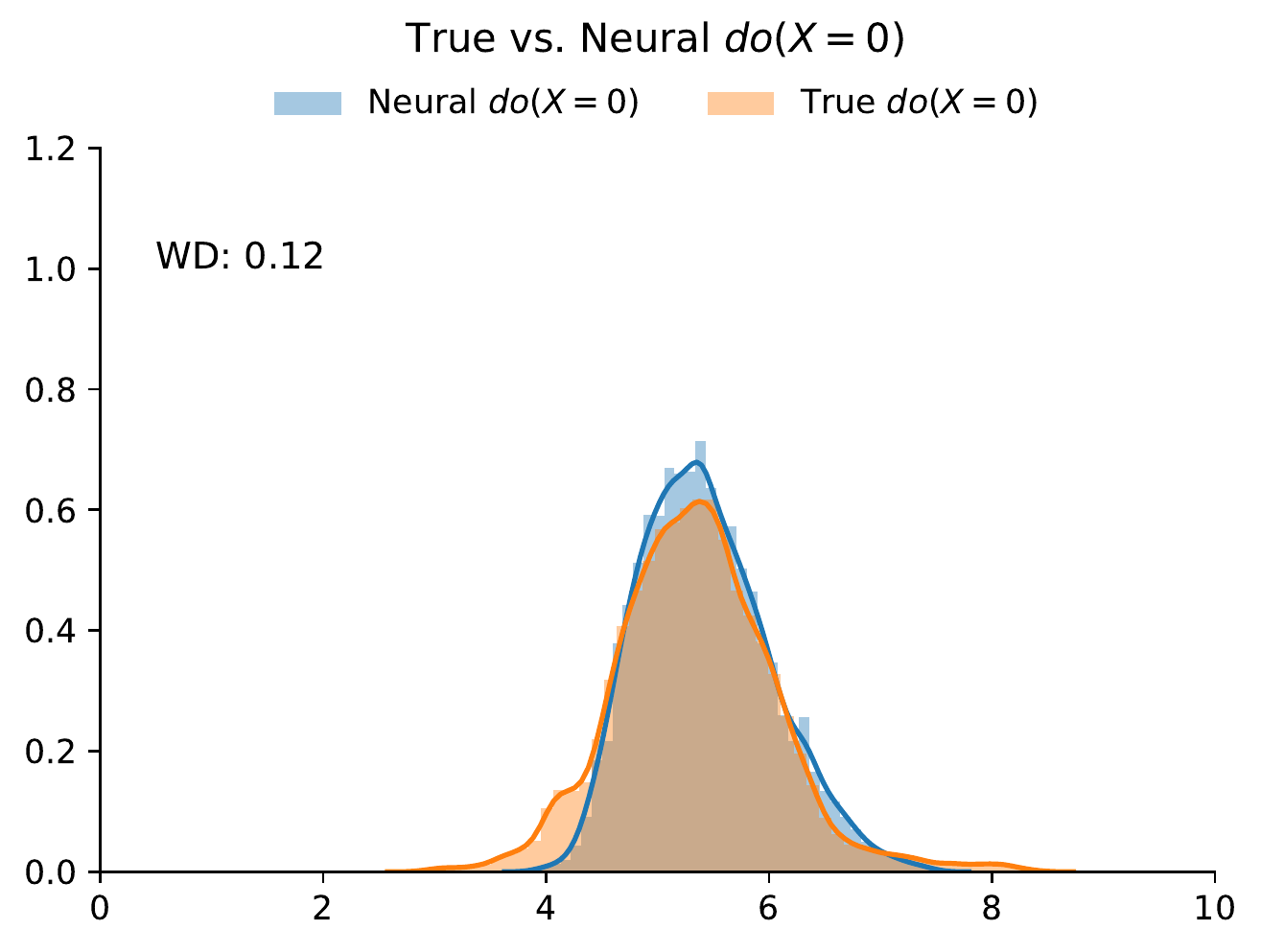}
  \includegraphics[width=0.3\textwidth]{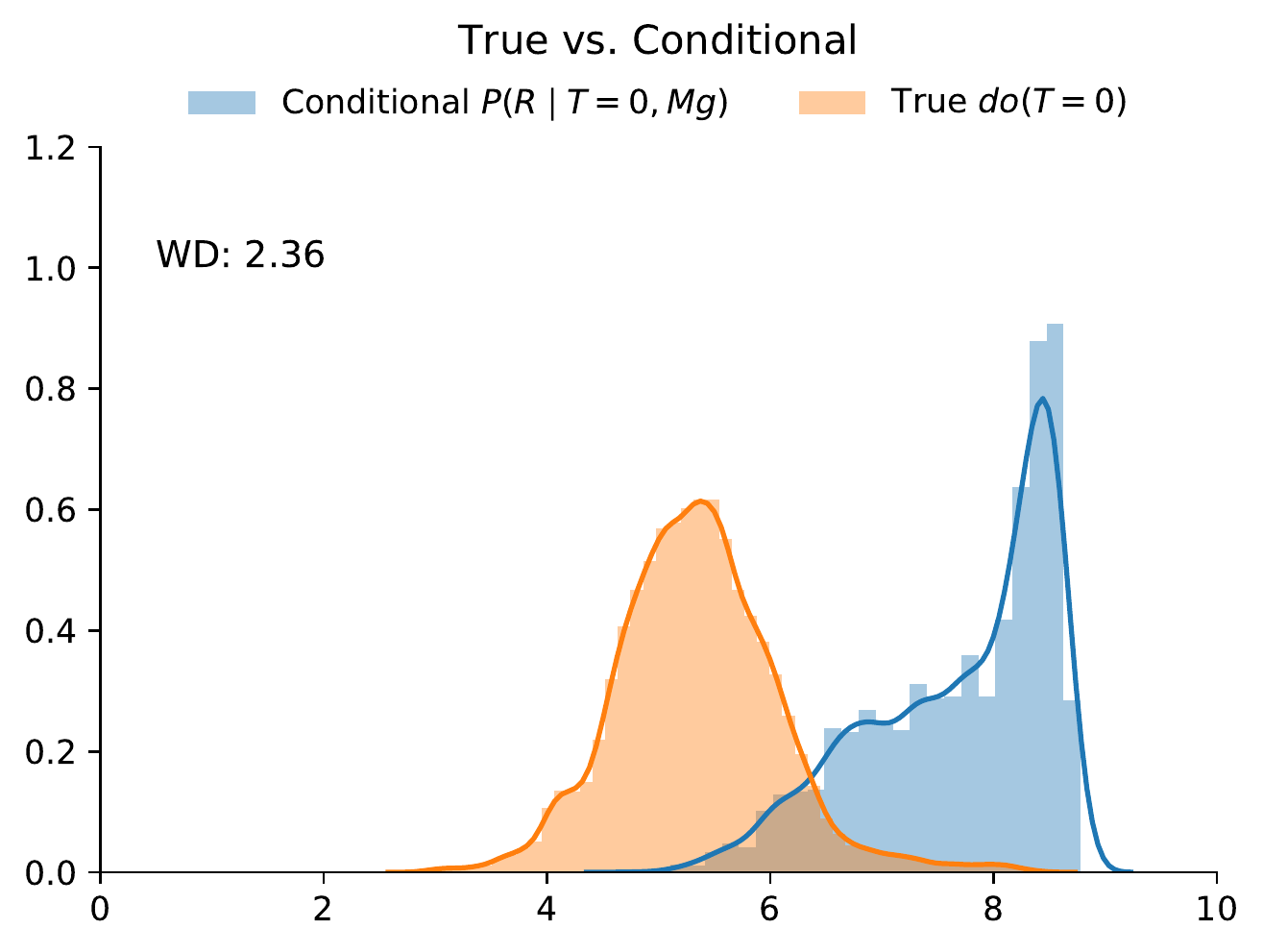} \\
  
  \includegraphics[width=0.3\textwidth]{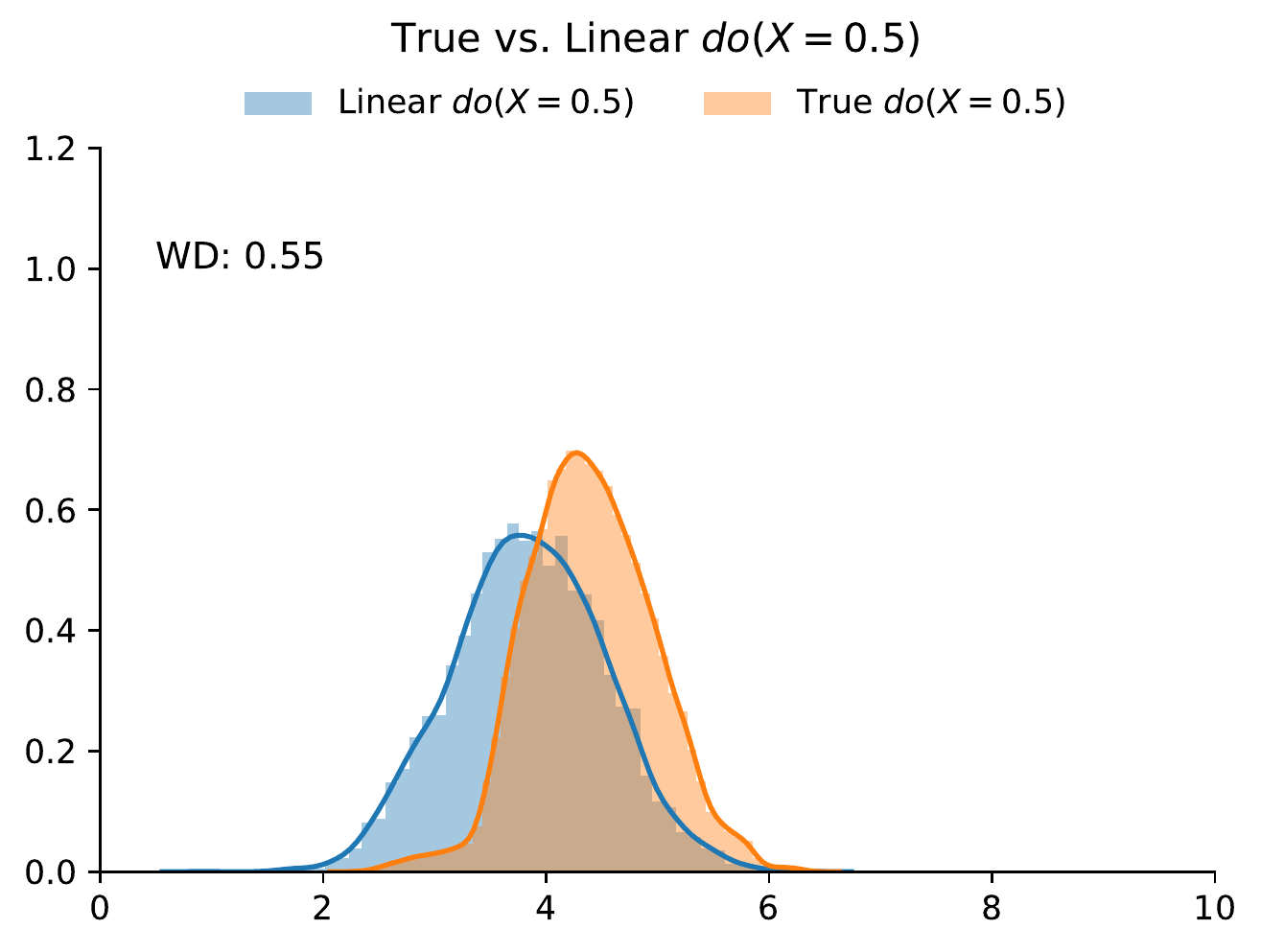}
  \includegraphics[width=0.3\textwidth]{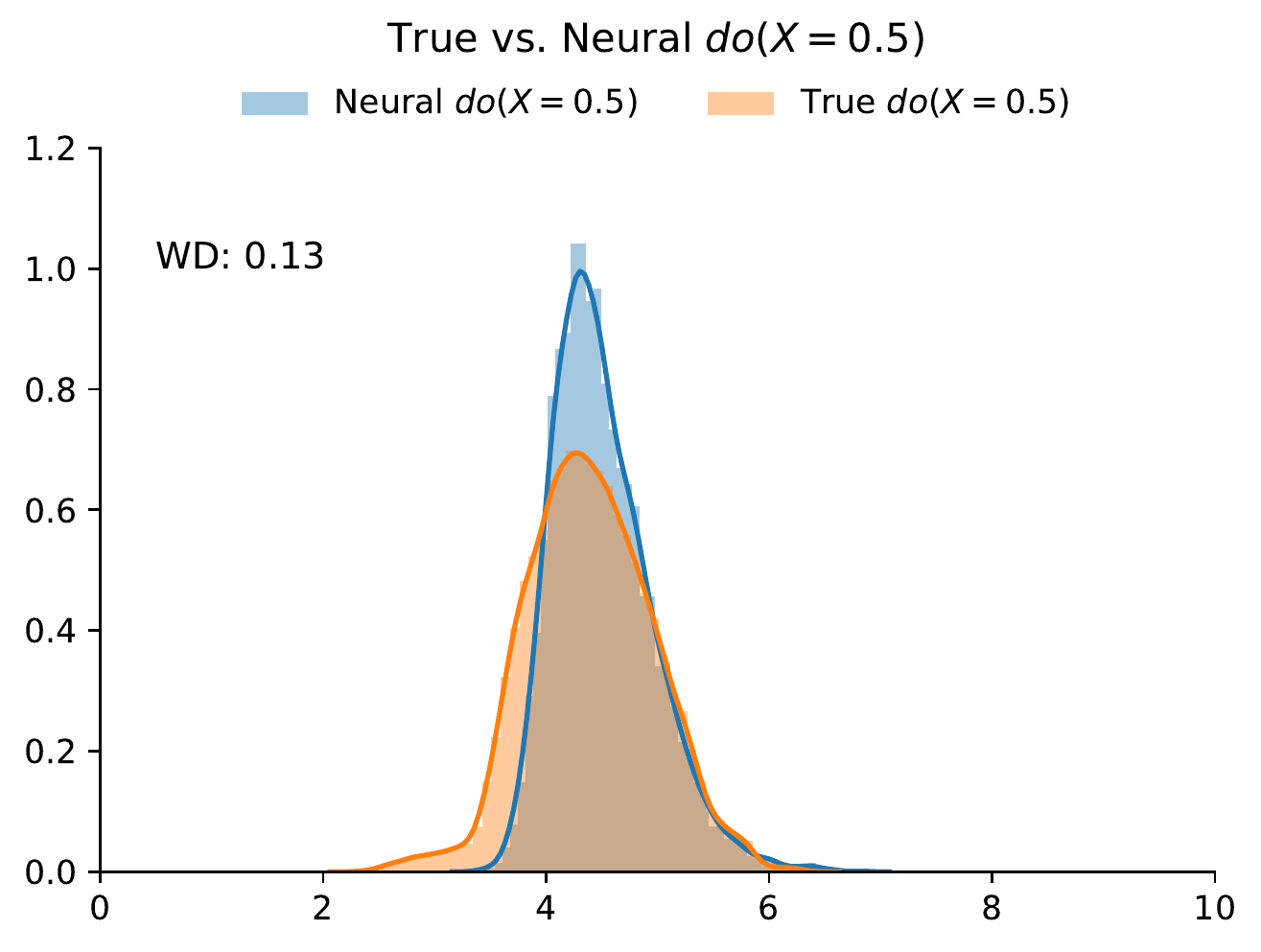}
  \includegraphics[width=0.3\textwidth]{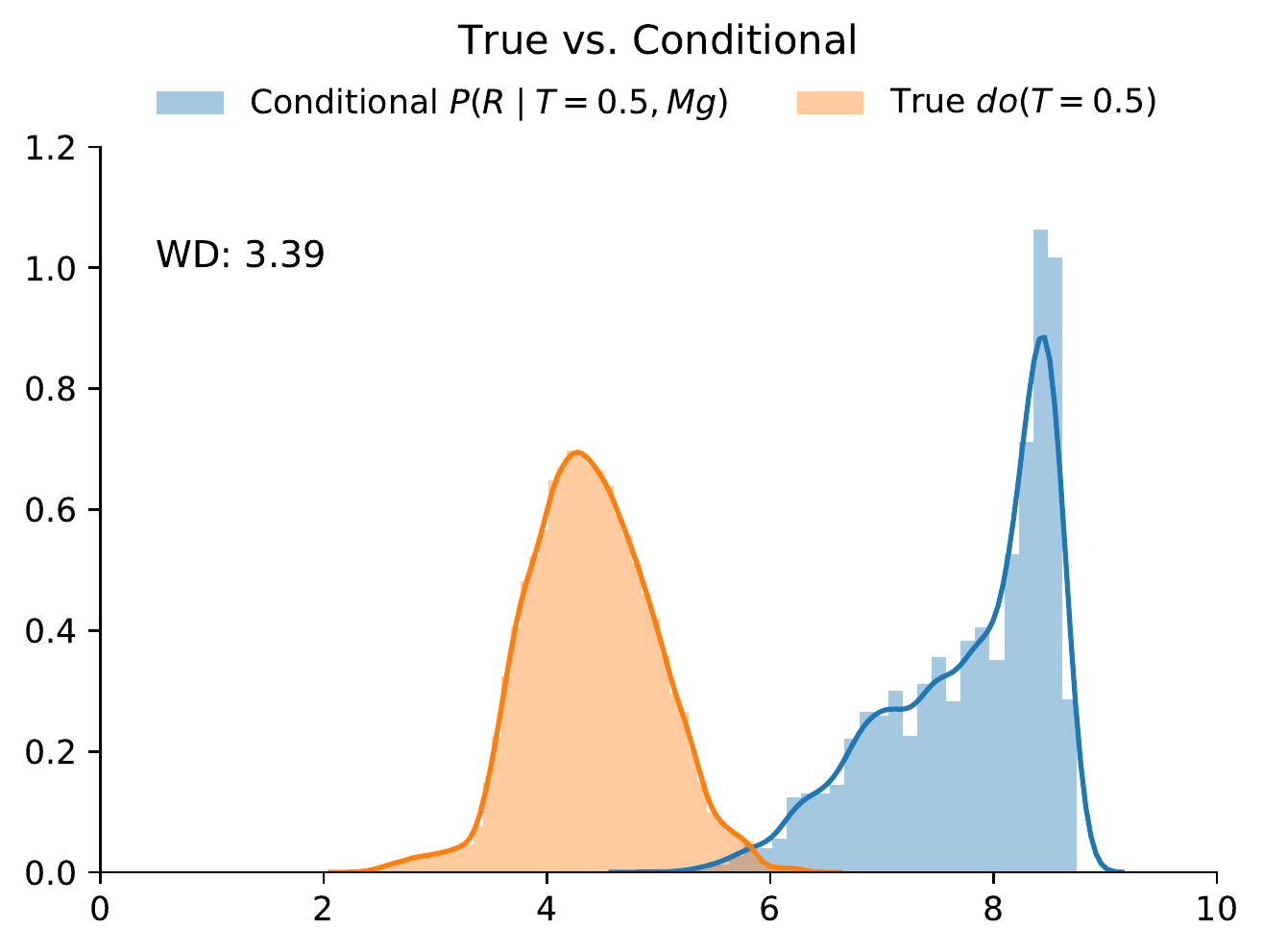}
  
  \caption{Comparison of three different ways to estimate causal effects. (Left) Linear model. (Center) Our proposed approach. (Right) Simple conditioning. (Above) Comparing against the true $do(X=0)$, and (Below) against the true $do(X=0.5)$. The Wasserstein Distance (WD) between the distributions is specified on each plot.}
  \label{fig:front-door-comparison}
\end{figure*}

In Figure~\ref{fig:front-door-comparison} we compare the estimated distribution of $R$ for three models with the respective true distribution. On the left, the linear model finds a single distribution for the different values of $T$, as expected. In the middle, our distribution is seen to match the true simulated distribution. The discrepancies between the true distribution and the real distribution arise from the different elements of stochasticity, e.g. the sampling of the true effects, the stochastic optimization procedure and the Monte Carlo approximation of the integral. In the plot to the right of Figure~(\ref{fig:front-door-comparison}) it is observed that pure conditioning with respect to $T$ and integrating over $Mg$ from the conditional network $P(R \mid X=x, Mg=mg)$ gives results that are far off the true distribution. We also included the Wasserstein Distance (WD) in every plot as a metric of comparison between the different models and the true distribution. The proposed approach performs better for the interventional distributions when compared to the other approaches.

\subsection{Experiment VI: Unobserved confounder}
\label{ssec: unobserved_confounder_experiment}

In this section we explore the robustness of our model when the assumption of knowing the causal graph is not fulfilled. To examine this, three additional experiments were carried out. Two experiments where the unobserved confounder has a linear relationship with the outcome variable with different ``degrees'' of confounding, and one experiment where the relationship is non-linear. In all cases, it was assumed that there is a single unobserved confounder $U$, which affects both treatment and outcome variables. The factor multiplying this term in the ``mild'' case is 0.3, and 3 for the ``strong'' case. The true causal graph and DGP can be summarized as follows:

\begin{figure*}[!htb]
    \centering
    \includegraphics[width=0.3\textwidth]{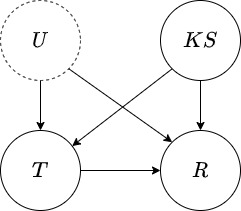}
    \caption{Causal graph for the unobserved confounder experiment}
    \label{fig:unobserved_confounder_causal_graph}
\end{figure*}

\begin{algorithm}
    \caption{DGP for unobserved confounder experiment}
    \label{alg:unobserved_confounder}
    \begin{algorithmic}[1]
        \FOR{each individual $i$}
        \STATE Draw $KS \sim$ Log-Normal$(2.5, 0.25)$, we subtract the mean of $KS$.
        \STATE Draw $U \sim$ Normal$(0, 1)$.
        \STATE Draw $T$ with $P(T=A) =  \frac{1}{1+exp((-KS-U)/10)}$.
        \STATE Draw $R \sim N \big(\frac{50T}{KS+3} + \textbf{0.3}TU, 1\big)$.
        \ENDFOR
    \end{algorithmic}
\end{algorithm}

\begin{figure*}[!htb]
  \includegraphics[width=0.49\textwidth]{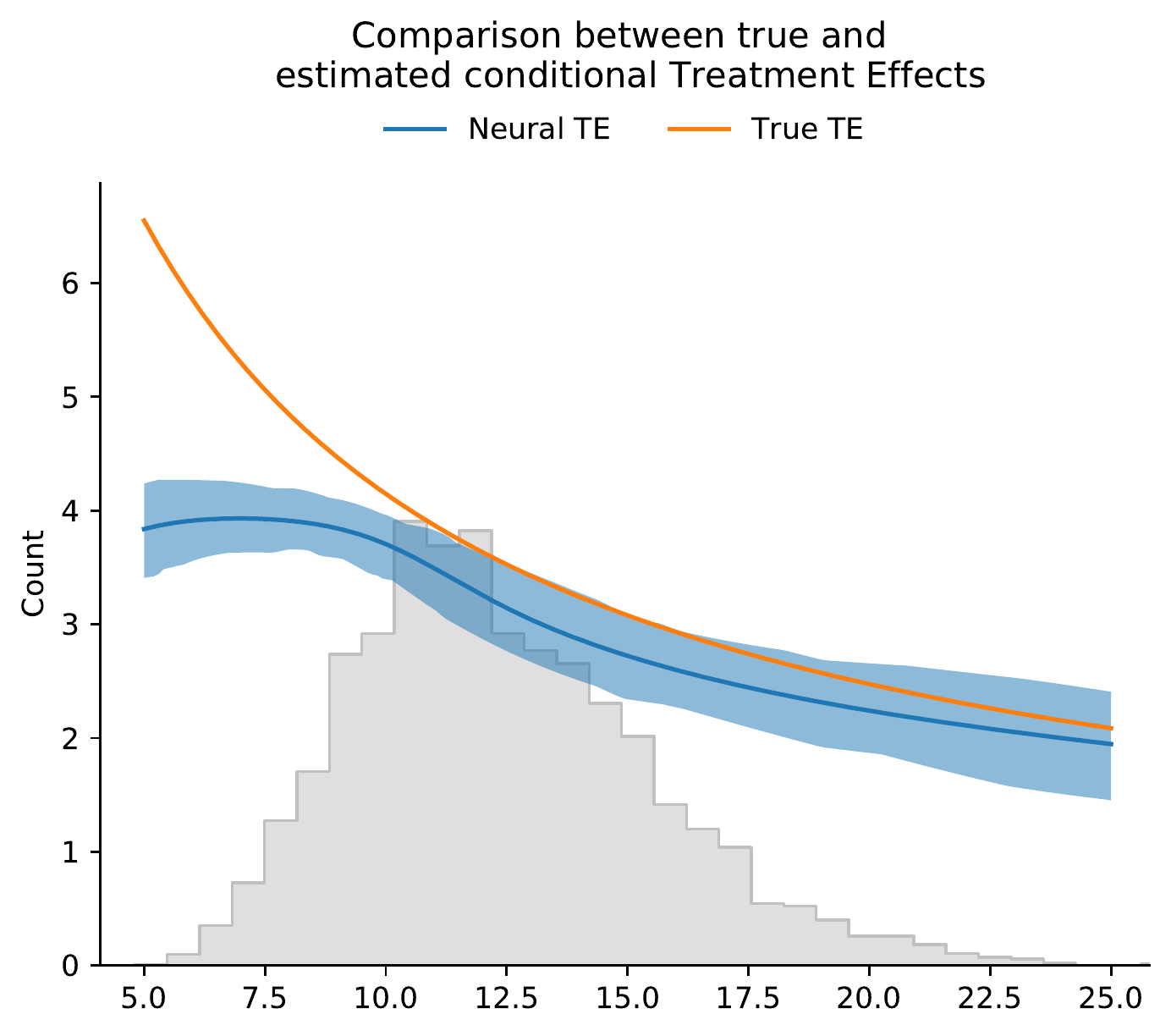}
  \includegraphics[width=0.49\textwidth]{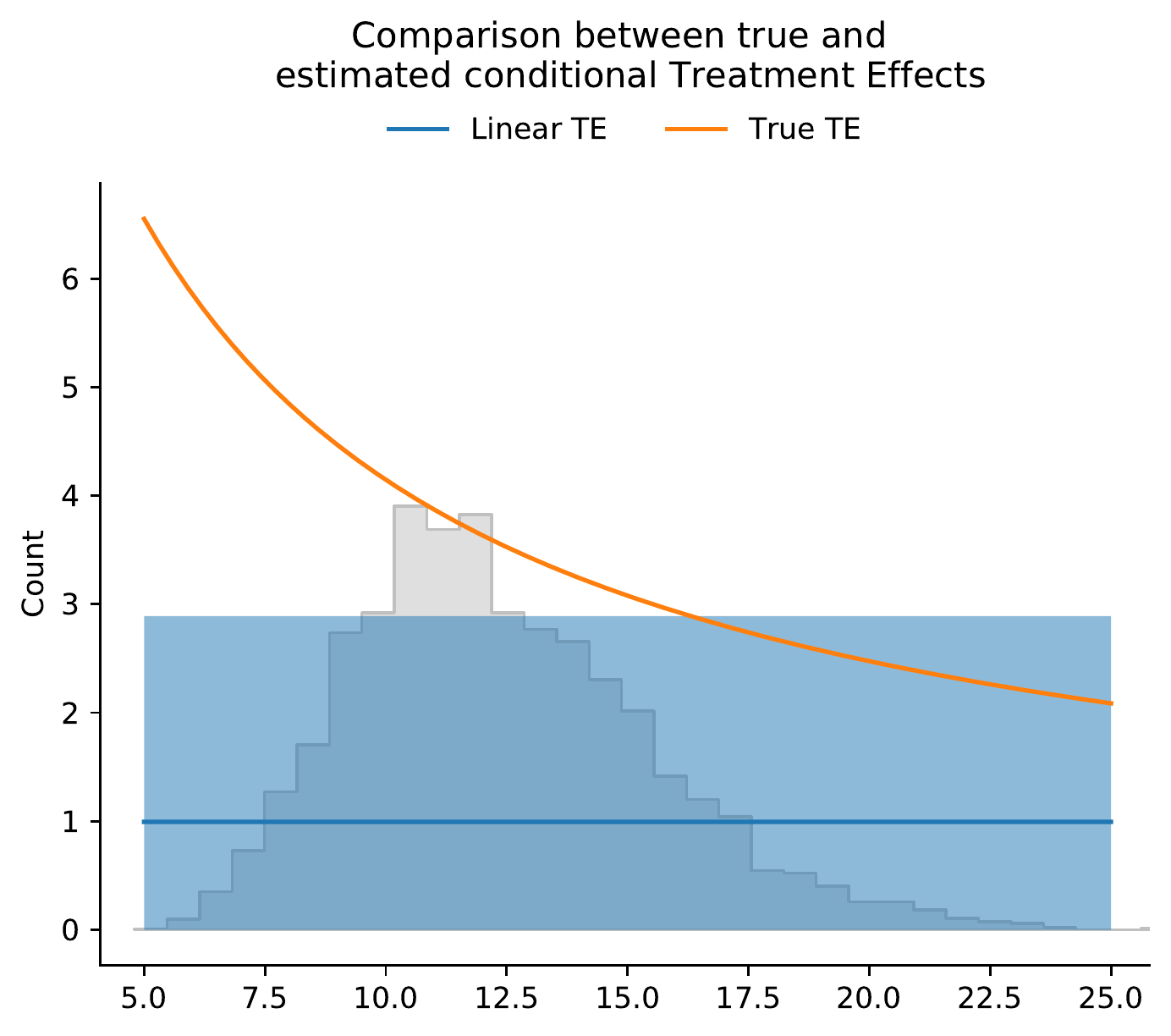}
  \caption{Comparison between the linear and neural network models using the back-door adjustment formula in the \textbf{mild} unobserved confounding scenario. The solid blue line represents the mean of the bootstrap, while the bands around the mean represent the 90\% confidence bands. A histogram of $KS$ is represented in the background of the plots.}
  \label{fig:bootstrap_mild_unobserved_confounding}
\end{figure*}

\begin{figure*}[!htb]
  \includegraphics[width=0.49\textwidth]{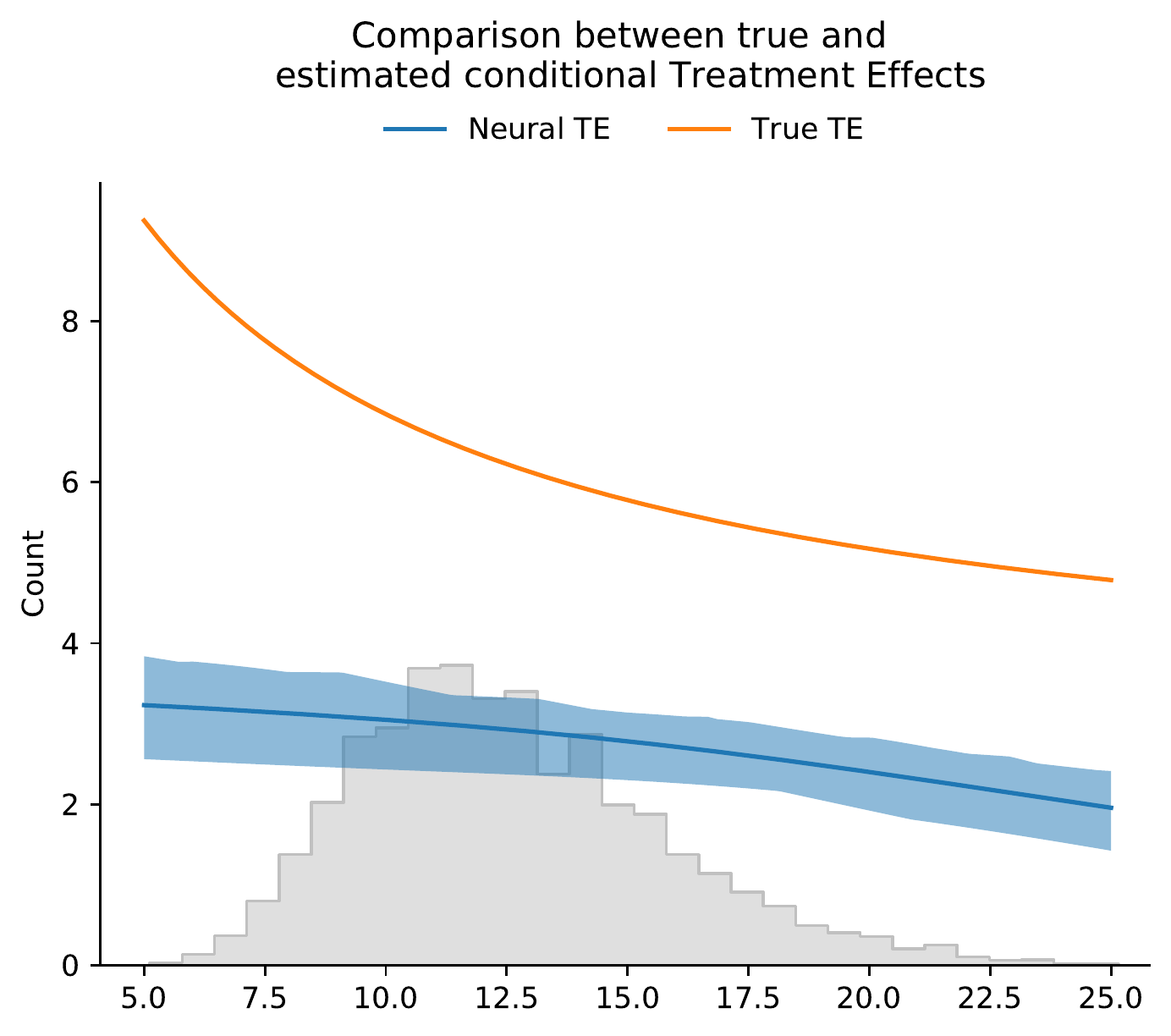}
  \includegraphics[width=0.49\textwidth]{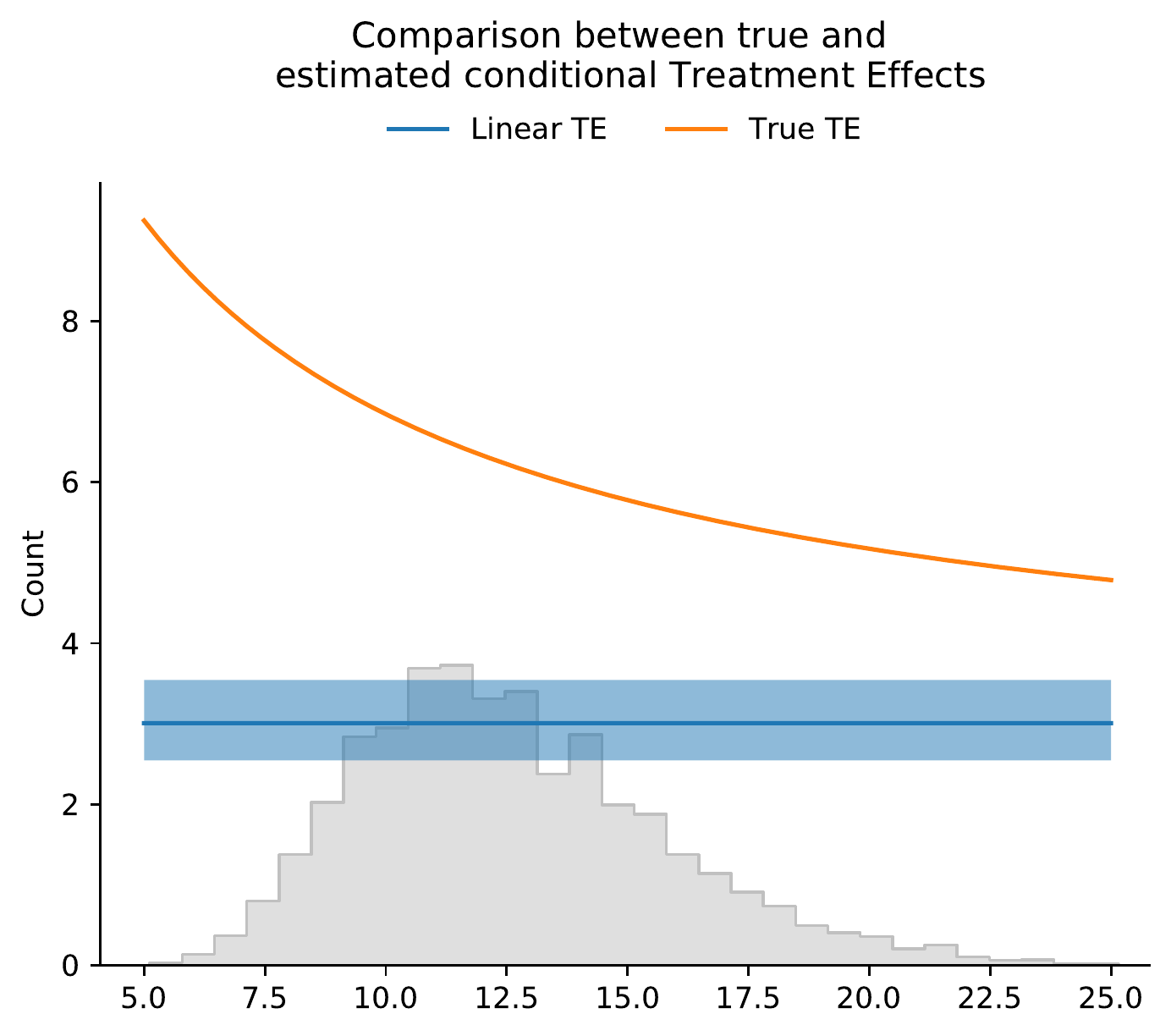}
  \caption{Comparison between the linear and neural network models using the back-door adjustment formula in the \textbf{strong} unobserved confounding scenario. The solid blue line represents the mean of the bootstrap, while the bands around the mean represent the 90\% confidence bands. A histogram of $KS$ is represented in the background of the plots.}
  \label{fig:bootstrap_strong_unobserved_confounding}
\end{figure*}

The assumption of a known graph structure is violated by ignoring the existence of $U$. In other words, the causal effect is estimated using the same architecture as in Case IV, using the data from Algorithm~\ref{alg:unobserved_confounder}. The results of the best performing models can be found in Figures~\ref{fig:bootstrap_mild_unobserved_confounding} and \ref{fig:bootstrap_strong_unobserved_confounding}.

Both figures show a deviation from the true treatment effect as expected. The results for both the neural and the linear model are relatively similar to those without unobserved confounding, in the sense that the neural model learns a non linear treatment effect and the confidence bands are relatively tight. Likewise, for the mild confounding case, the linear model learns a linear treatment effect with big confidence bands. It is interesting, however, that in the strong confounding case, the confidence bands of the linear model are considerably tighter than the rest of the models. This might be due to the fact that the confounding captures most of the variance of the outcome variable and, consequently, the best fit linear model is the one closest to the value of the confounding.

The second type of model with which we tested the assumption of a known causal graph is a model where the relationship between the unobserved confounder and the outcome variable is non-linear. The following DGP is used to simulate the data:

\begin{algorithm}
    \caption{DGP for the unobserved confounder experiment, including non-linearity}
    \label{alg:nonlinear_unobserved_confounder}
    \begin{algorithmic}[1]
        \FOR{each individual $i$}
        \STATE Draw $KS \sim$ Log-Normal$(2.5, 0.25)$, we subtract the mean of $KS$.
        \STATE Draw $U \sim$ Normal$(0, 1)$.
        \STATE Draw $T$ with $P(T=A) =  \frac{1}{1+exp((-KS-U)/10)}$.
        \STATE Draw $R \sim N \big(\frac{50T}{KS+3} + TU^{2}, 1\big)$.
        \ENDFOR
    \end{algorithmic}
\end{algorithm}

\begin{figure*}[!htb]
  \includegraphics[width=0.49\textwidth]{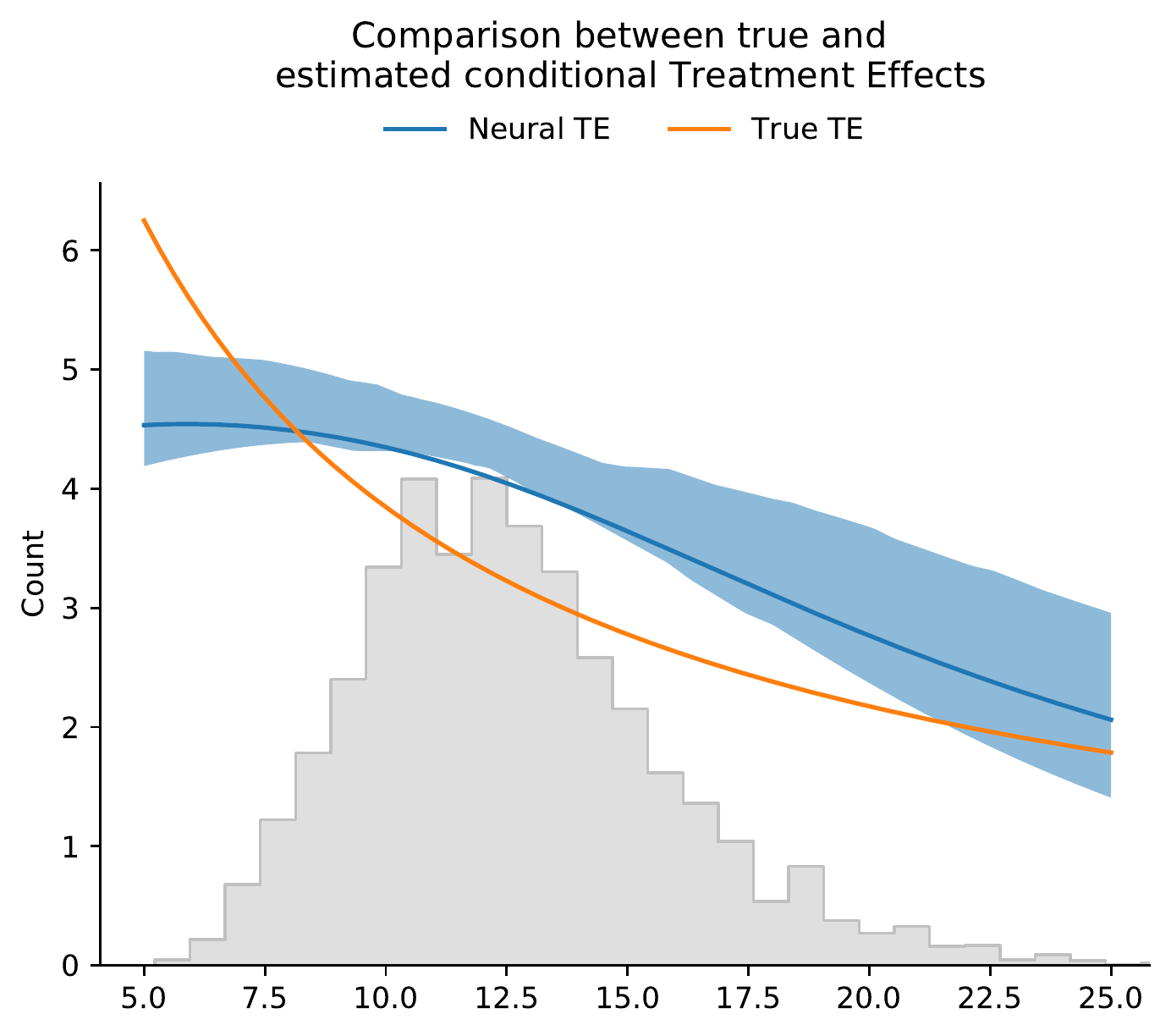}
  \includegraphics[width=0.49\textwidth]{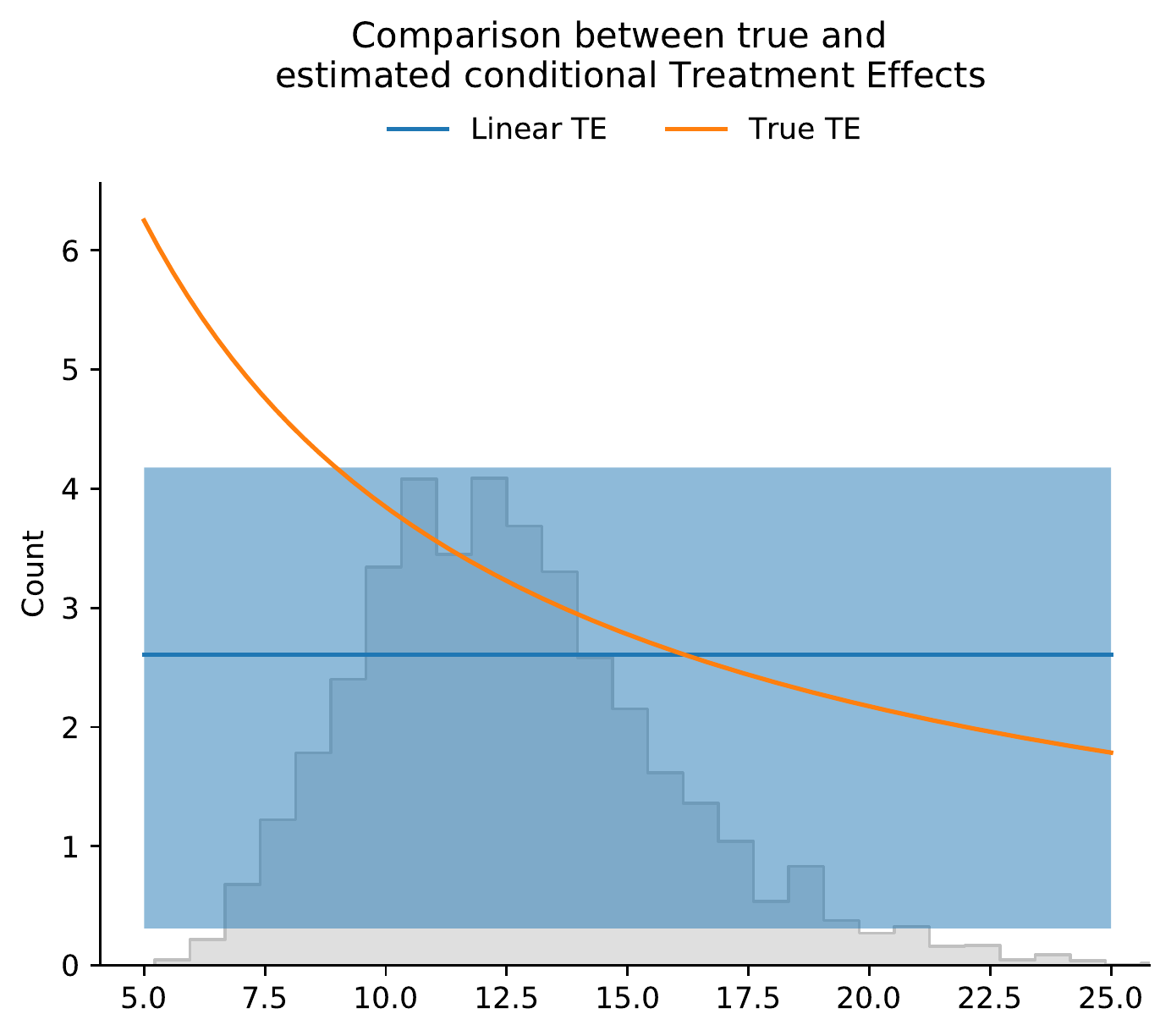}
  \caption{Comparison between the linear and neural network models using the back-door adjustment formula in the \textbf{non-linear} unobserved confounding scenario. The solid blue line represents the mean of the bootstrap, while the bands around the mean represent the 90\% confidence bands. A histogram of $KS$ is represented in the background of the plots.}
  \label{fig:bootstrap_nonlinear_unobserved_confounder}
\end{figure*}

The estimated causal effects using this simulation is presented in Figure~\ref{fig:bootstrap_nonlinear_unobserved_confounder}.

As expected from all the simulations where the causal graph is not completely known, the estimated causal effects are far from the actual effects regardless of whether the estimator is linear or non-linear. The main lesson of this study cases is that, regardless of the estimator being used, if the true causal graph is not known, the estimators are not going to be particular strong predictors for the outcome. 


\section{Conclusions and future work}

In this paper, we propose the neural autoregressive density estimator to estimate the causal conditionals. The motivation of this is to avoid having to make any unnecessary assumptions with respect to the functional form of the independent causal mechanisms. We show in several simulation examples, with different specifications, that the proposed method is able to recover both linear and non-linear relations, superseding any linear model. 

Several conclusions are offered. It is shown that our estimator is effective in recovering non-linear functions when the causal graph is known and the model is supported by data. Based on simulation and bootstrapping, we explore the robustness of the estimator when we diverge from the assumption of knowing the causal graph and having strong data support. As expected, both of these assumptions affect the model results. However, even with limited data support the model provides a reasonable good approximation, as long as the causal graph is known. Assumptions concerning knowing the causal graph are found to be more critical for the model performance. Hence, not adjusting for unobserved confounders will decrease the quality of the estimated causal effects, regardless of the flexibility of the estimation method.

As a final contribution, the paper can be seen as a bridge between theoretical methods from the causality literature and advanced methods from the statistical and machine learning literature.

The paper points to several new research directions. Machine learning researchers who commonly apply neural networks as a tool to capture complex unknown relationships can introduce additional structure to their models by integrating causal models into their framework. Such causal models may go beyond the traditional domains in the causal literature, and consider settings with data coming from images, audio, and speech. On the other hand, methods from the machine learning literature could also deem useful to social sciences in situations where the systems are well studied (in a causal way) but where functional relationships are not known exactly.

In future research, it would be interesting to explore the potential uses of the learned independent causal mechanisms (or single parent-child networks). For example, they could be used to transfer learning (also known as transportability in the causality literature \citep{pearl_transportability_2014}), or continual learning. In the same context, it would be interesting to study the learned representations in the independent causal mechanisms and whether these representations could be used for ``downstream" tasks, for example, in reinforcement learning \citep{zhang2020designing}.

\appendix
\section{Detailed experimental setup}
\label{apx:experimental_setup}

For every experiment, we selected the best performing hyperparameters for three type of neural network activations: linear (which collapses to a linear regression), ii) Rectified Linear Units (ReLU), and iii) hyperbolic tangent (Tanh). For each one of them, two different optimizers: RMSProp and Stochastic Gradient Descent were tested. For each of the optimizers. four different learning rates were used: $1e^{-2}$, $5e^{-3}$, $1e^{-3}$, and $5e^{-4}$. Finally, five different architectures for the neural networks were tested: 1 Hidden Layer (HL) with 4 units, 1 HL with 8 units, 1 HL with 16 units, 2 HL with 4 units each, and 2 HL with 8 units each. For this paper, a total of $9x3x4x2x5=1080$ neural networks were trained. All the neural networks were trained on the Python programming language, using the PyTorch \cite{paszke2017automatic} package for neural network estimation and Numpy \cite{harris2020array} for synthetic data generation.

We selected a single neural network for each experiment and each activation function based on its performance with respect to the ground truth using the Mean Absolute Error (MAE) metric. Alternative metrics such as the Root Mean Squared Error (RMSE) could also be used, however, the results would be similar. After the best architecture was selected, we ran 50 bootstrap samples in order to build the 90\% confidence intervals shown in the plots.

Interesting observations were made during the experimental stage of the research. \textbf{First}, the ReLU activation function was inferior in all experiments to both the tanh and linear activation functions, in the sense that all ReLU architectures had greater error than any tanh or linear architecture. \textbf{Second}, the estimates of the interventional distribution under the front-door adjustment are highly variable. We hypothesize this is the case, because the number of Monte Carlo samples used to approximate the second integral of Formula~\ref{eq:frontdoor_adjustment} is only one. \textbf{Third}, no single combination of hyperparameters seemed to be superior to others. This is a challenge as in real case scenarios as the ground truth causal effects are unobserved (otherwise there is no causal effect learning task). Nevertheless, it was found that better performing networks usually have higher likelihood with respect to the training data compared to those networks that performed poorly. This was not the case in all experiments but there is a correlation between high likelihood and the models that estimate the causal effects closer to the ground truth. This is not a trivial observation, as all the networks were trained to maximize the likelihood: why should the models with higher likelihood should also estimate the causal effects best? 





\newpage
\nocite{*}

\bibliographystyle{apalike}
\bibliography{causal_nade.bib}

\end{document}